\documentclass[%
 reprint,
 amsmath,amssymb,
 aps,
prl,
]{revtex4-1}

\usepackage{graphicx}
\usepackage{dcolumn}
\usepackage{bm}

\begin{document}

\preprint{APS/123-QED}


%
%
%

\title{Quantum Oscillations, Thermoelectric Coefficients, and the Fermi Surface of Semimetallic WTe$_{2}$}
\author{Zengwei Zhu$^{1,*}$, Xiao Lin$^{2}$, Juan Liu$^{1}$, Beno\^{\i}t Fauqu\'{e}$^{2}$, Qian Tao$^{2}$, Chongli Yang$^{3}$, Youguo Shi$^{3}$, and Kamran Behnia$^{2}$}

\affiliation{(1)Wuhan National High Magnetic Field Center, School of Physics, Huazhong University of Science and Technology,  Wuhan  430074, China\\
(2)LPEM (CNRS-UPMC), ESPCI, 75005 Paris, France\\
(3)Institute of Physics and Beijing National Laboratory for Condensed Matter Physics, Chinese Academy of Sciences, Beijing 100190, China\\}

\date{\today}

\begin{abstract}
 We present a study of angle-resolved quantum oscillations of electric and thermoelectric transport coefficients in semi-metallic WTe$_{2}$, which has the particularity of displaying a large B$^{2}$ magneto-resistance. The Fermi surface consists of two pairs of electron-like and hole-like pockets of equal volumes in a ``Russian doll'' structure. Carrier density, Fermi energy, mobility and the mean-free-path of the system are quantified. An additional frequency is observed above a threshold field and attributed to magnetic breakdown across two orbits. In contrast to all other dilute metals, the Nernst signal remains linear in magnetic field even in the high-field ($\omega_c\tau \gg 1$) regime. Surprisingly, none of the pockets extend across the c-axis of the first Brillouin zone, making the system a three-dimensional metal with moderate anisotropy in Fermi velocity yet a large anisotropy in mean-free-path.

\begin{description}
\item[PACS numbers]71.18.+y, 72.15.Jf, 71.20.Gj
\end{description}
\end{abstract}

\pacs{Valid PACS appear here}
\maketitle



Manyfold change in electric conduction induced by magnetic field is dubbed giant\cite{GMR} and colossal\cite{CMR} magnetoresistance. The spin degree of freedom plays an important role in both cases. On the other hand, by coupling  to the charge degree of freedom through the Lorentz Force, magnetic field can enhance the resistivity of a solid hosting extremely mobile carriers by many orders of magnitude. As early as 1928, Kapitza \cite{kapitza1928} discovered that this orbital magnetoresistance is very large in archetypal semi-metals such as bismuth and graphite.

Recently, large magnetoresistance in dilute metals such as Cd$_{3}$As$_{2}$ \cite{cd2As3}, NbSb$_{2}$\cite{WangKF} and WTe$_{2}$\cite{Ali} has attracted attention. In the particular case of WTe$_{2}$, a quadratic magnetoresistance (i.e proportional to B$^{2}$) was reported with no sign of saturation up to 60 T\cite{Ali}. This is the expected behavior of a perfectly-compensated semi-metal\cite{Pippard1989}, but has not been seen in  bismuth\cite{SatBi} or graphite\cite{SatGra}, two compensated semi-metals whose Fermi surface is accurately known. A first step towards uncovering the ultimate reason behind the quadratic magnetoresistance of WTe$_{2}$ is a quantitative determination of the structure of the Fermi surface and the components of the mobility tensor.

In this letter, we report on a study of quantum oscillations of resistivity, Seebeck and Nernst coefficients in high-quality single crystals of  WTe$_{2}$ and find that the Fermi surface consists of two pairs of electron-like and hole-like pockets. Each pair is concentric with identical structure like a set of Russian dolls. The anisotropy is much smaller than one would naively expect in a layered system. The longer axis of the pockets is much shorter than the height of the Brillouin zone, in contrast to the theoretical expectations. Moreover, we find another distinctive feature of this semi-metal in addition to quadratic magnetoresistance, which is a Nernst response linear in magnetic field deep inside the high-field limit. Our results quantify carrier concentration of the system and set plausible quantitative windows for mobilities and Fermi energies leading to the huge quadratic magnetoresistance and large field-linear Nernst signal.

\begin{table}[ht]\label{Fermi}
\caption{The Residual Resistivity Ratio (RRR),  the c-axis magnetoresistance at 10 T [MR= $\frac{\Delta\rho(10T)} {\rho(0)}$] and the average mobility, $\overline{\mu}^{2}=\frac{\Delta\rho}{B^{2}\rho(0)}$ of the WTe$_{2}$ sample used in this study compared to those previously reported.} 
\centering 
\begin{tabular}{c c c c} 
\hline\hline 
 Sample & RRR   & $ MR (10 ^{3})$ & $\overline{\mu}(T^{-1})$ \\ [0.5ex] 
\hline 
This work & 1256 & 31& 17.6 \\
Ali \emph{et al.}\cite{Ali} & 370 & 2.1 & 4.6 \\
Cai \emph{et al.}\cite{shiyan}& 184 & 0.6 & 2.4 \\
Zhao \emph{et al.}\cite{Zhao}& 741 & 5.2 & 7.2 \\
\hline\hline
\end{tabular}
\label{table:nonlin} 
\end{table}

Ribbon-like single crystals with typical dimensions of 2$\times$0.3$\times$0.1mm$^{3}$  were prepared by a solid state reaction by using Te as the flux (more details in SM\cite{SM}), the longest along the a-axis, which was also the orientation of electric/thermal current in all our measurements. Nernst and Seebeck coefficients were measured by a standard one-heater-two-thermometers configuration in a dilution refrigerator and  samples were rotated in two perpendicular planes using piezoelectric rotators as previously\cite{Zhu1,Zhu2}. Density-Functional-Theory band calculations were performed using the WIEN2K code (see the SM section for details\cite{SM}).

\begin{figure}
\includegraphics[width=7cm]{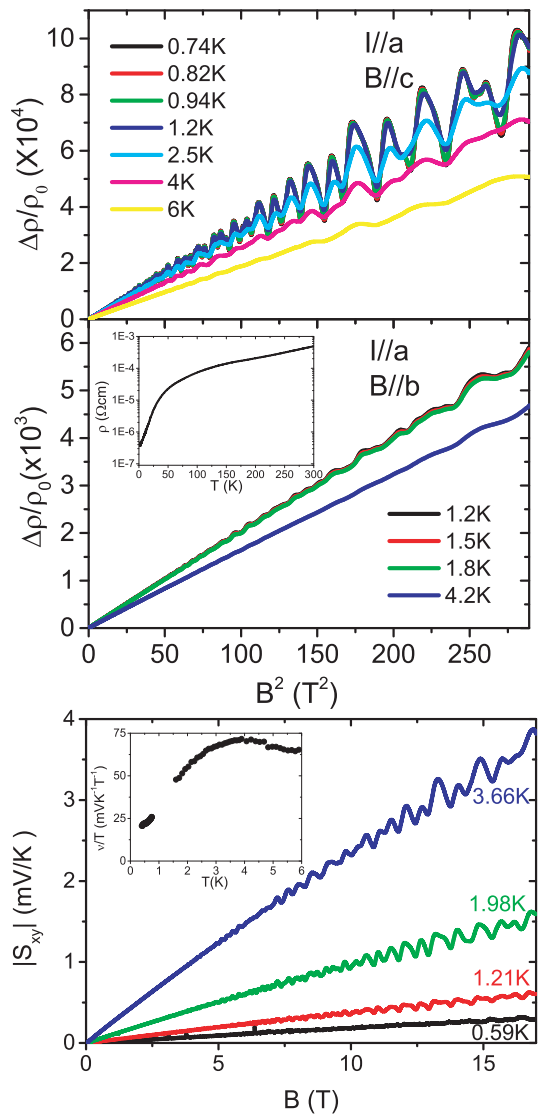}
\caption{(Color online) Transverse magnetoresistance for magnetic field along c-axis (panel a) and along b-axis (panel b) for different temperatures plotted as a function of B$^{2}$. The inset shows the temperatures dependence of resistivity at zero magnetic field.  Shubnikov-de Haas oscillations emerge at high magnetic field. For both orientations, magnetoresistance is quadratic, but the B$^{2}$ slope is much larger for B//c.  (c). The large Nernst signal is linear in magnetic field up to 17 T. The inset shows the temperature dependence of $\nu/T$, which reveals a large diffusive component at zero temperature limit.  }
\end{figure}

The temperature dependence of resistivity in absence of magnetic field and magnetoresistance for two orientations of magnetic field perpendicular to the current are shown in two panels and the inset of Fig.1. Like previous studies, we find that transverse magnetoresistance follows a B-square behavior for both B//c and B//b configurations. As seen in table I, the ratio of room-temperature resistivity (490$\mu\Omega$cm) to residual resistivity (0.39$\mu\Omega$cm) in our sample is significantly larger than those studied previously\cite{Ali,shiyan,Zhao}. Unsurprisingly, orbital magnetoresistance is also much larger in our sample. Assuming a simple scalar mobility $\overline{\mu}$,  orbital magnetoresistance would be $\frac{\Delta\rho(B)}{\rho(0)}= \overline{\mu}^{2} B^2$. As we will see in more detail below, the correlation seen in the table between RRR and $\overline{\mu}$ points to a quasi-isotropy of the mobility tensor in the (a, b) plane. On the other hand, the reduced slope of the B-square magnetoresistance for the B//b configuration (see  Fig. 1b), points to a much lower mobility along the c-axis.

Fig. 1c  presents the field dependence of the Nernst signal at different temperatures and reveals another peculiarity of WTe$_{2}$. The magnitude of the Nernst coefficient, $\nu$, is large, as one may expect in a system with a low concentration of mobile carriers\cite{Kamran}. As seen in the inset, in addition to the large phonon drag contribution,  a diffusive $\nu/T$ as large as 20$\mu$VK$^{-2}$T$^{-1}$ can be resolved in the zero-temperature limit. This is lower than bismuth, but larger than graphite\cite{Zhu2010}. In contrast to these two semi-metals, however, the Nernst signal shows no visible deviation from linearity up to 17 T. This feature  puts WTe$_{2}$, at odd, not only with other semi-metals, but also with all other semiconductors turned to dilute metals by doping, such as Bi$_{2}$Se$_{3}$\cite{Fauque},  SrTiO$_{3-\delta}$\cite{Lin} and Pb$_{1-x}$Sn$_{x}$Se\cite{Liang}. In all these systems, a downward deviation from linearity emerges when one attains the strong-field limit ($\mu B >1$ ) and quantum oscillations emerge. Liang \emph{et al.}\cite{Liang} showed that such a deviation at the cross-over between low-field and high-field limits is expected in the semiclassical picture. The argument requires a number of conditions (See the supplement). According to our finding, they are not fulfilled in the case of WTe$_{2}$. Thus, among dilute metals WTe$_{2}$ stands out by a quadratic magnetoresistance and a field-linear Nernst signal, both pointing to the specificity of its mobility tensor components. An accurate determination of the Fermi surface is the prelude to quantify these components.

\begin{figure}
\includegraphics[width=9cm]{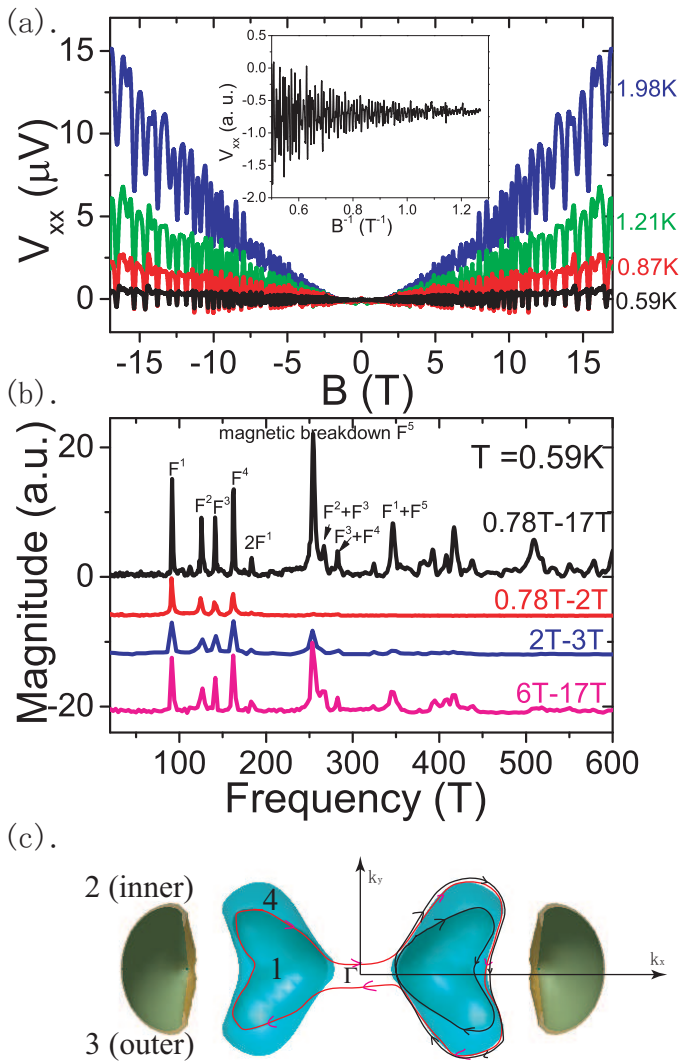}
\caption{(Color online) a)Quantum oscillations of the Seebeck response at several temperatures are huge on top of a very small background. Inset shows the data at T = 0.59 K as a function of  B$^{-1}$. b) FFT analysis of the quantum oscillations of the Seebeck coefficient. At the low-field below 2T, the 250T peak is absent. But the magnitude of this frequency increases gradually with increasing magnetic field, and finally dominates FFT spectrum.  c) The calculated Fermi surface with two possible breakdown paths marked by red and black. Two hole pockets and electron pockets are labelled as 1, 2, 3 and 4.}
\end{figure}

Quantum oscillations, visible in magnetoresistance and Nernst data of Fig. 1, were also observed in longitudinal thermoelectric (Seebeck) response.  The results for B//c are shown in Fig. 2. As seen in the inset, oscillations become visible as soon as the magnetic field exceeds 0.8 T, another testimony to the extreme  mobility of carriers in this sample. Thanks to the large thermoelectric oscillations studied in high-quality crystals, a fine picture of the Fermi surface can be drawn and the main parameters are summarized in Table II.

\begin{table}[ht]\label{Fermi}
\caption{Quantum-oscillation frequencies, $F_{a,b,c}$, for three orientations of magnetic field along the high symmetry axes, the extracted Fermi radius, k$_{a,b,c}$ of the four pockets assimilated to a triaxial ellipsoid Fermi surface and the carrier concentration of each pocket expressed in units of 10$^{19}$cm$^{-3}$.}
\centering 
\begin{tabular}{c c c c c c c c}
\hline\hline
 & F$_{c}(T)$& F$_{b}(T)$ & F$_{a}(T)$ & k$_{a}$(nm$^{-1}$) & k$_{b}$ (nm$^{-1}$) & k$_{c}$(nm$^{-1}$) &n \\ [0.5ex]
\hline 
1(h) &  92 & 192 & 233 & 0.48 & 0.59 &1.22& 1.17\\ 
2(e) & 125 & 208 & 280 & 0.536 & 0.71 &1.19& 1.53\\
3(e) & 142 & 220 & 319 & 0.545 & 0.791 &1.23& 1.79\\
4(h) & 162 & 267 & 375 & 0.594 & 0.831 &1.37& 2.28\\ [1ex] 
\hline 
\end{tabular}
\label{table:nonlin} 
\end{table}

We can identify four distinct intrinsic frequencies corresponding to two pairs of electrons and hole pockets. The type of carrier for the frequencies  Our band calculations produce results similar to those previously reported\cite{Ali, Zhao} and were used to identify the pockets. The theoretical Fermi surface (Fig. 2c) consists of two small hole pockets, slightly off the $\Gamma$ point and sandwiched by two electron pockets along the $\Gamma$-X direction in the Brillouin zone\cite{ThoeryWTe2}. They are larger than what we find by experiment in agreement with the previous  ARPES study\cite{ARPESWte2}. As we will see below, there is also an important difference in the fine structure of the pockets.

In addition to the four fundamental frequencies, a fifth frequency (F$_{c}^{5}$ = 254T) is also detected. Because of its sudden appearance above 2T, we attribute it to magnetic breakdown across two pockets. This hypothesis is backed by the fact that F$_{c}^{5}$ is equal to the sum of F$_{c}^{1}$ and F$_{c}^{4}$. This makes it likely that it corresponds to a high-field orbit enclosing two adjacent low-field orbits. Since this frequency dominates the FFT spectrum above 2T, it is unlikely to be a second harmonic of one of the basic frequencies as suggested by a previous study\cite{shiyan}. Two possible paths for magnetic breakdown are sketched in the figure and are to be contrasted with the case of $\kappa$-(BEDT-TTF)$_{2}$Cu(NCS)$_{2}$\cite{MBNeil}. Note that the Fermi pockets might be closer to each other in the momentum space than what is expected from band calculations. Recent APRES measurements found significant quasi-particle weight at $\Gamma$ point\cite{FengARPES}.  If the pockets on either sides of the $\Gamma$-point happen to be almost touching each other, it would provide a natural explanation for the unusually robust magnetic breakdown observed here.

\begin{figure}
\includegraphics[width=9cm]{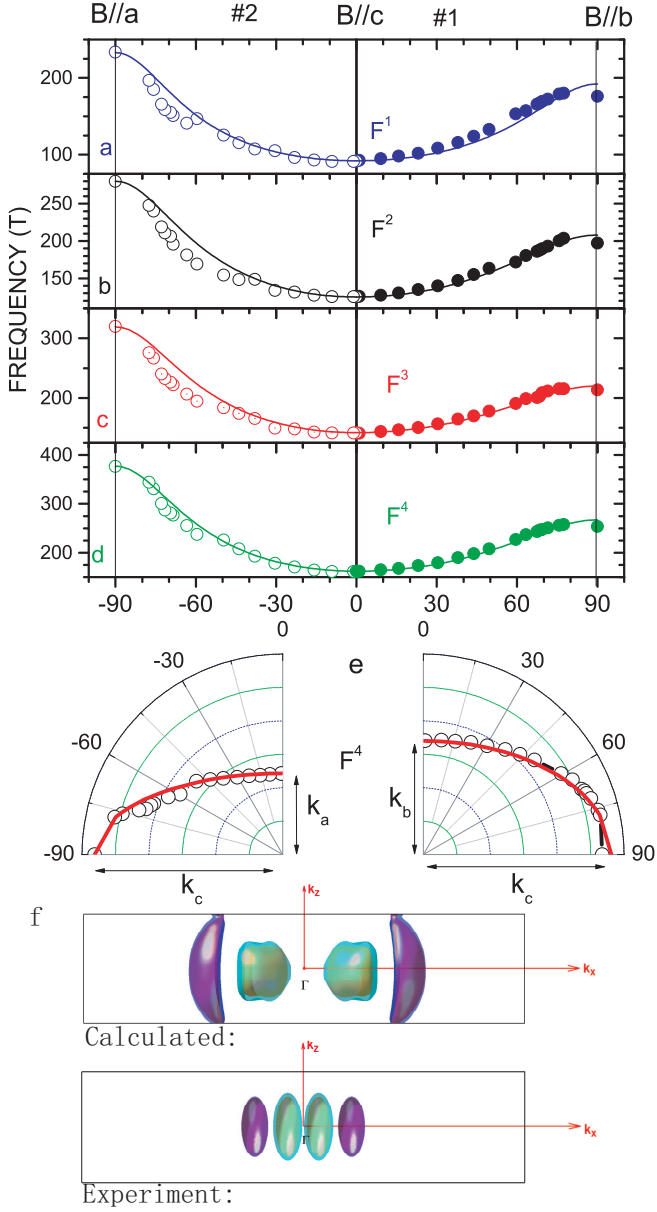}
\caption{(Color online) a-d) Angular evolution of the FFT frequencies extracted from quantum oscillations of the Seebeck coefficient for a field rotating in (a, c) and (a, b) planes in samples  \#1 and  \#2 (symbols) compared to what is expected for a triaxial ellipsoid (solid lines). e) Comparison of the effective radius extracted from our data (symbols) to the radius of the perfect ellipsoid (solid lines), in the case of pocket 4.f) Theoretical and experimental Fermi surfaces inside the Brillouin zone projected to the (k$_{x}$, k$_{z}$) plane.Note that our experiment cannot quantify the distance between the pockets, but the occurrence of magnetic breakdown at a very low magnetic field suggests an extreme proximity between the two inner pockets. The locations
of Fermi pockets might be different from the figure.}
\end{figure}

A detailed study of the evolution of the quantum oscillations with the orientation of magnetic field  allowed us to map the fine structure of the Fermi surface. Two samples \#1 and \#2 were used to rotate the magnetic field in both (a, c) and  and (b, c) planes. The evolution of extracted frequencies in the two planes are shown in Fig.3. Treating each pocket as a triaxial ellipsoid with three distinct semi-axis lengths,  k$_{a}$,k$_{b}$ and k$_{c}$, we used the Onsager relation ($F=(\hbar/2\pi e)A_{k}$ between frequency, F, and the extreme cross section, A$_{k}$, of a Fermi surface) to extract the values listed in Table II. Based on the theoretical band structure, pockets 1 and 4 (2 and 3) can be identified as hole-like(electron-like). As seen in Fig. 3, the data shows clear deviation from what is expected for a perfect ellipsoid, which is not surprising, given what is theoretically expected. However, these deviations are not large. As seen in Fig.3e, the effective radius(see SM also about the fitting of an ellipsoid Fermi surface.\cite{SM}) deviates from what is expected for a perfect ellipsoid by 15 percent over a limited angular range.  No difference in morphology among the pockets is visible, which is not surprising if they are two pairs of concentric ``Russian dolls''.

The structure of the Fermi surface have multiple quantitative consequences. The first is the concentration of carriers. Ellipsoid volumes yield the carrier density for each electron-like or hole-like pocket and are listed in table II. Keeping in mind that there are two sets of such pockets in the Brillouin zone,  the total hole and electron density is n= 6.64 $\times$ 10$^{19}$cm$^{-3}$ and p= 6.9 $\times$ 10$^{19}$cm$^{-3}$. Thus,  WTe$_{2}$ is  a compensated metal within a precision of 4 percent, comparable to what has been achieved in the case of bismuth\cite{Bhargava1967}.

Second, the anisotropy of the Fermi surface is surprisingly  mild, given the layered structure of WTe$_{2}$.  For all four pockets, the k$_{c}$/k$_{a}$ ratio is between two and three.  In particular, in contrast to what is theoretically expected, 2k$_{c}< 2.8nm^{-1}$ along k$_{z}$, is shorter than the height of the Brillouin zone along the c-axis ($\frac{2\pi}{c}=4.5nm^{-1}$), implying three-dimensionality. The Fig.3f shows the comparison of the experimental and theoretical Fermi surface projected to the (k$_{x}$, k$_{z}$) plane of the Brillouin zone. In comparison, in graphite the hole and electron ellipsoids lie on top of each other, extend from  the bottom to the top of the Brillouin zone  and their k$_{c}$/k$_{a}$ ratio is as large as nine for holes and seven for electrons\cite{graAni}.

Third, it gives a picture of the anisotropy of mobility and the mean-free-path. By combining carrier density and residual resistivity and using $\rho_{0}^{-1}=(n+p)e\mu_{a}$, one finds a mobility along the a-axis ($\mu_a=12 T^{-1}$). Plugging this to the magnitude of quadratic magnetoresistance (for B//b and B//c) yields $\mu_b=26 T^{-1}$ and $\mu_c=1.8 T^{-1}$)(See SM\cite{SM}). The large mobility along b-axis implies a mean-free-path as long as 12 $\mu$m (using  $\mu=\frac{e\ell}{\hbar k_{F}}$, remarkably long, but still an order of magnitude shorter than the longest reported for Cd$_{3}$As$_{2}$\cite{cd2As3}). On the other hand, along the c-axis, the mean-free-path is one order of magnitude shorter indicating that carriers traveling across the planes are severely scattered.

Fourth, the temperature dependence of quantum oscillations were used to quantify the cyclotron mass, in the range of 0.3-0.8m$_{e}$ and the Fermi energy of E$_{F} \sim$20-40 meV. Combined to the measured mobility, this gives a fair account of the magnitude of the diffusive Nernst coefficient. The threshold field at which the magnetic breakdown occurs (B$^{*}$= 2T),  and the energy gap between the breakdown orbits can be linked together through $\hbar\omega_{c}\gtrsim\varepsilon_{g}^{2}/E_{F}$, \cite{Shoenberg}, indicating an $\varepsilon_{g}$ as small as a few meV which is consistent with theory(See SM for more theoretical estimation\cite{SM}).

Strict equality between concentrations of electrons and holes is a necessary, but not sufficient condition for non-saturating B$^{2}$ magneto-resistance. Elemental bismuth is believed to be perfectly compensated\cite{Liu1995}. Its complex angular magnetoresistance can be quantitatively described by semi-classic theory\cite{Zhu2012,Collaudin2015}. Nevertheless, the field dependence of magnetoresistance is not quadratic for any orientation of magnetic field. This is because the carrier concentration and the mobility tensor change with magnetic field.  WTe$_{2}$ is different, because the carrier density is such that the quantum limit is far away and no field-induced change in carrier density is expected.

In summary, we mapped the Fermi surface of WTe$_{2}$, and found two hole-like and two electron-like pockets with similar morphology and a compensation within four percent. This leads to a quantitative description of the mobility, the magnetoresistance, the Nernst signal and the threshold field for magnetic breakdown. We also found that in addition to quadratic magnetoresistance, a field-linear Nernst response distinguishes WTe$_{2}$ from other dilute metals.

Z. Z. acknowledges Yuke Li's help in the initial stage of this research and thanks Luo Bo for the transparent visual processing of Fermi surface is supported by the 1000 Youth Talents Plan, and 111 program. J. L is supported by the National Natural Science Foundation of China under Grant No. 11404121. Y. S. is supported by the ¡°Strategic Priority Research Program(B)¡± of the Chinese Academy of Sciences, Grant No. XDB07020100. The work in France is supported by ANR through the SUPERFIELD program.

* \verb|zengwei.zhu@hust.edu.cn|

\end{document}